\documentclass[fleqn,10pt]{wlscirep}
\usepackage[utf8]{inputenc}
\usepackage[T1]{fontenc}
\usepackage{subcaption}
\usepackage{float}

\title{A crowdsensing intrusion detection dataset for decentralized federated learning models}

\author[1,*]{Chao Feng}
\author[1,2]{Alberto Huertas Celdran}
\author[1]{Jing Han}
\author[1]{Heqing Ren}
\author[1]{Xi Cheng}
\author[1]{Zien Zeng}
\author[1]{Lucas Krauter}
\author[3]{Gérôme Bovet}
\author[1]{Burkhard Stiller}
\affil[1]{Communication Systems Group, Department of Informatics, University of Zurich UZH, 8050 Zürich, Switzerland}
\affil[2]{Department of Information and Communications Engineering, University of Murcia, 30100 Murcia, Spain}
\affil[3]{Cyber-Defence Campus, armasuisse Science \& Technology, 3602 Thun, Switzerland}
\affil[*]{cfeng@ifi.uzh.ch}


\begin{abstract}
This paper introduces a dataset and an experimental study on Decentralized Federated Learning (DFL) for Internet of Things (IoT) crowdsensing malware detection. The dataset comprises behavioral records from benign and eight malware attacks. A total of 21,582,484 original records were collected from system calls, file system activities, resource usage, kernel events, input/output events, and network records. These records were aggregated into 30-second windows, resulting in 342,106 data records used for model training and evaluation. Experiments on the DFL platform compare traditional Machine Learning (ML), Centralized Federated Learning (CFL), and DFL across different node counts, topologies, and data distributions. Results show that DFL maintains competitive performance while preserving data locality, outperforming CFL in most settings. This dataset provides a solid foundation for studying the security of IoT crowdsensing environments.
\end{abstract}
\begin{document}

\flushbottom
\maketitle
\thispagestyle{empty}

% \noindent Please note: Abbreviations should be introduced at the first mention in the main text – no abbreviations lists. Suggested structure of main text (not enforced) is provided below.

\section*{Background \& Summary}
The Internet of Things (IoT) has permeated nearly every aspect of the physical world, from smart homes to industrial automation~\cite{huertas2022intelligent}. These devices enable a wide variety of sensing, actuation, and automation applications, but their large-scale deployment also introduces new challenges for monitoring, managing, and securing such heterogeneous and resource-constrained networks~\cite{celdran2022privacy}.

One particularly relevant IoT application scenario is crowdsensing, where a large number of distributed, heterogeneous devices collaboratively collect and contribute measurements or observations about their local environments~\cite{rajendran2017electrosense}. Crowdsensing leverages the collective sensing capability of many participants to enable large-scale monitoring without relying on a centralized infrastructure. This paradigm is appealing for applications such as traffic monitoring, environmental sensing, and anomaly detection, especially in privacy-sensitive and bandwidth-constrained settings. However, the distributed and open nature of crowdsensing networks also makes them susceptible to security threats~\cite{celdran2022privacy}.

In particular, crowdsensing devices are often targeted by malware and intrusion attacks, which undermine the availability, integrity, and functionality of the system~\cite{huertas2022intelligent}. Detecting such intrusions and identifying compromised devices has therefore become a critical research challenge, especially in large-scale, dynamic crowdsensing deployments. Machine Learning (ML)-based anomaly detection methods have been widely adopted for this task, as they can analyze multidimensional behavioral data from devices to identify deviations indicative of compromise. However, conventional ML pipelines typically assume that all behavioral data can be transmitted to a centralized server for training, which raises privacy concerns and creates vulnerabilities by concentrating data and computation in a single location~\cite{beltran2023decentralized}.

Federated Learning (FL) has emerged as a privacy-preserving alternative to centralized ML for collaborative intrusion detection~\cite{nguyen2019diot}. In FL, each device retains its local data and performs on-device training, sharing only model updates with a coordinating server that aggregates them into a global model, called Centralized FL (CFL). This architecture mitigates privacy risks while benefiting from distributed knowledge~\cite{rey2022federated}. Nevertheless, the reliance on a central server introduces a single point of failure, potential scalability bottlenecks, and vulnerability to targeted attacks, limitations that are particularly pronounced in large, dynamic crowdsensing scenarios.

To address these issues, Decentralized FL (DFL) has been proposed. By removing the client-server hierarchy and adopting peer-to-peer topologies, DFL enables each node to act both as a learner and as an aggregator~\cite{beltran2023decentralized}. This fully decentralized design aligns naturally with the characteristics of crowdsensing, enhancing scalability, robustness, and privacy in collaborative intrusion detection for IoT environments.

However, despite the growing interest in DFL as a promising framework for IoT intrusion detection, the lack of suitable benchmark datasets has become a key impediment to research progress. Several benchmark datasets have been developed for intrusion detection research, most notably in the context of centralized ML. The NSL-KDD dataset~\cite{tavallaee2009detailed} is a widely used successor to the KDD’99 benchmark, providing labeled network connection records. However, it is outdated and does not reflect modern IoT-specific traffic or attack patterns. The CICIDS2017 dataset~\cite{sharafaldin2018toward} addresses some of these limitations by offering flow-based network traffic data with a wider variety of attacks, but it still assumes a centralized data collection scenario and lacks device-level heterogeneity. To specifically target IoT environments, the Bot-IoT dataset~\cite{koroniotis2019} was proposed, simulating IoT device traffic mixed with botnet attacks in a testbed environment. While Bot-IoT captures some IoT-specific characteristics, its synthetic nature and centrally collected data limit its realism for decentralized learning evaluations. Similarly, the TON\_IoT dataset~\cite{alsaedi2020ton_iot} provides telemetry, operating system, and network data from IoT and industrial control systems. However, the setup of TON\_IoT involves homogeneous devices and centrally collected data, lacking client-level heterogeneity and device-specific data distributions that are essential for evaluating DFL in realistic IoT settings.

These limitations highlight the need for a realistic, distributed, and well-documented dataset specifically tailored to evaluating DFL frameworks in crowdsensing-based IoT intrusion detection tasks. Such a dataset would enable systematic research on privacy-preserving, scalable, and robust anomaly detection methods under realistic deployment conditions.

To address the lack of datasets and benchmarks for evaluating DFL in crowdsensing-based IoT intrusion detection, a dataset was constructed and validated through experimental evaluation. Therefore, the main contributions of this study are as follows:
(1) the behavioral dimensions of IoT devices affected by malware attacks were analyzed, identifying network activity, input/output operations, file system access, resource usage, system calls, and kernel events as key data to monitor. This informed the design of the experimental platform and the selection of features;
(2) an experimental setup consisting of six Raspberry Pi 3 and two Raspberry Pi 4 devices connected to the ElectroSense platform~\cite{rajendran2017electrosense} was implemented. Data were collected under eight malware attacks and a benign state, yielding 288 hours of monitoring and over 21,582,484 records. The dataset was cleaned, normalized, and processed to ensure suitability for machine learning while reducing overfitting risk;
(3) a DFL pipeline was developed within the Nebular framework~\cite{beltran2025nebula} to demonstrate the dataset’s applicability to DFL-based intrusion detection. A multilayer perceptron (MLP) model was used for malware classification. The DFL models were trained and evaluated under multiple DFL topologies, and their performance was measured using standard classification metrics, confirming the dataset’s utility for crowdsensing scenarios.

\section*{Methods}
\begin{figure}[h!]
    \centering
    \includegraphics[width=0.68\linewidth]{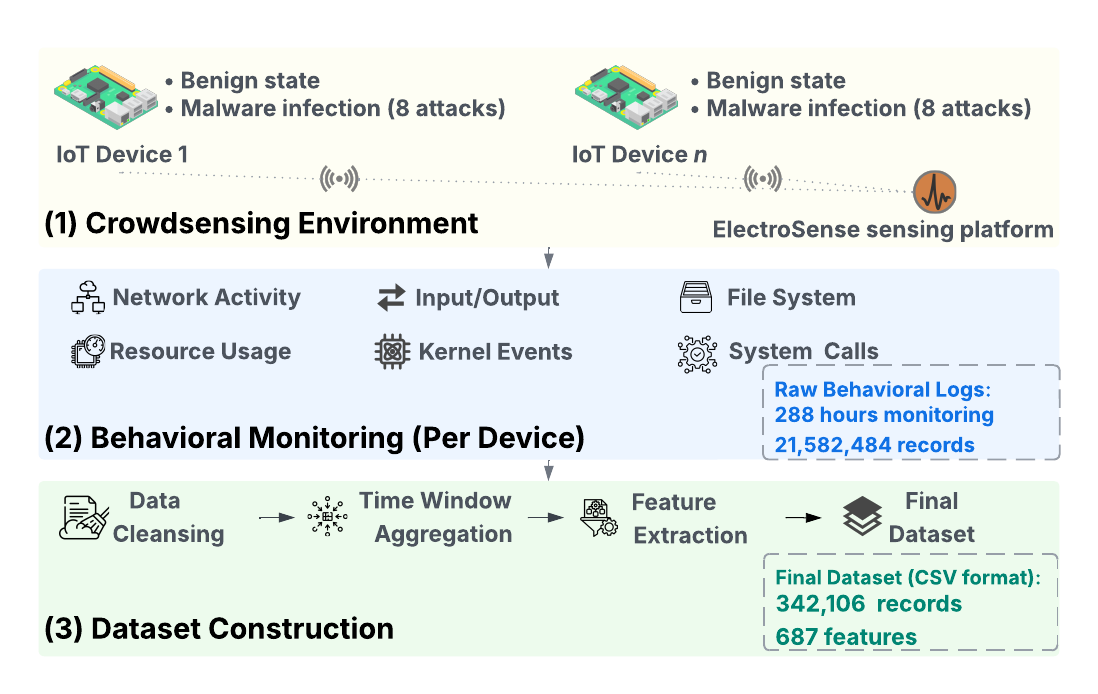}
    \caption{Workflow of the proposed dataset construction. IoT devices generate behavioral data under benign and malware-infected conditions, which are continuously monitored across six system dimensions. The resulting raw logs are cleansed, aggregated into fixed time windows, and converted into structured feature vectors, forming the final dataset.}
    \label{fig:workflow}
\end{figure}

This section describes the experimental design, data collection procedures, and data processing steps conducted in this work. The design of the experimental platform, the selection of behavioral dimensions to monitor, and the construction of the dataset are detailed to ensure reproducibility. The overall workflow of the proposed dataset construction and validation process is illustrated in \figurename~\ref{fig:workflow}. It shows the data collection at the device level, the subsequent data processing and feature engineering steps, and the DFL training and aggregation performed collaboratively by IoT devices. 

As shown in \figurename~\ref{fig:workflow}, the dataset construction process consists of three main stages. First, IoT devices operate in a crowdsensing environment under both benign and malware-infected conditions, generating behavioral data across multiple system dimensions. Second, each device monitors six behavioral sources and produces raw behavioral logs over an extended monitoring period. Finally, the collected logs are cleansed, aggregated into fixed time windows, and transformed into structured feature representations, resulting in a standardized CSV dataset suitable for ML and FL research.

\subsection*{Malware Families Affecting IoT Devices}
To ensure that the dataset captures diverse behavioral patterns representative of real-world IoT threats, five major malware families were selected: Botnets, Backdoors, Rootkits, Coinminers, and Ransomware. These families were chosen based on their prevalence in IoT environments and their distinct impacts on system-level behaviors~\cite{huertas2022intelligent}. Each family includes one or more open-source malware implementations executed on the experimental devices.

\textbf{Botnets.} Botnets consist of compromised devices remotely controlled by attackers to perform coordinated malicious activities such as distributed denial-of-service (DDoS) attacks, credential brute-forcing, or malware propagation. Botnet malware typically generates distinctive network traffic patterns, increased outbound connections, and frequent system call activity. In this dataset, Bashlite~\cite{bashlite} is used as a representative IoT botnet implementation.

\textbf{Backdoors.} Backdoors enable unauthorized remote access by bypassing standard authentication mechanisms. They allow attackers to execute commands, transfer files, or retrieve system information. Backdoor activity commonly affects system calls, file system operations, and network communication. Three backdoor implementations are included in the dataset: HttpBackdoor~\cite{httpbackdoor}, Backdoor~\cite{backdoor}, and TheTick~\cite{thetick}.

\textbf{Rootkits.} Rootkits are designed to conceal malicious presence by manipulating operating system components, often at the kernel level. They may hide processes, modify system utilities, or intercept kernel events. Rootkit behavior is typically reflected in file system operations, kernel events, and system call anomalies. The dataset includes two preload-based rootkits: Beurk~\cite{beurk} and Bdvl~\cite{bdvl}.

\textbf{Coinminers.} Coinminer malware exploits device computational resources to mine cryptocurrencies without user consent. Such malware primarily impacts CPU utilization, memory consumption, and I/O behavior, often while maintaining moderate network communication to avoid detection. XMRig~\cite{xmrig} is used as a representative coinminer in the dataset.

\textbf{Ransomware.} Ransomware encrypts files or system resources and demands payment for decryption. Its behavior strongly affects file system operations, input/output activity, and system calls related to file manipulation. The dataset includes Ransomware-PoC~\cite{ransomwarepoc}, an AES-based ransomware implementation executed under controlled conditions.

By including these five families and eight specific malware implementations, the dataset captures heterogeneous behavioral impacts across multiple system dimensions, supporting comprehensive evaluation of intrusion detection approaches in decentralized IoT settings.

\subsection*{Behavioral Selection}
The behavior of IoT devices under malware attacks is complex and multifaceted, involving a wide range of runtime activities and system interactions. Monitoring and recording all possible behaviors in their entirety is impractical due to the high overhead and limited relevance of many behaviors to malware detection. Therefore, it is necessary to select a subset of behavioral dimensions that are most effective for identifying malware. 

\begin{table}[b]
\centering
\caption{Behavioral dimensions affected by different malware types in IoT environments.}
\label{tab:malware_behavior}
\scalebox{0.85}{
\renewcommand{\arraystretch}{1.2}
\begin{tabular}{lp{2.5cm}p{2.5cm}p{2.5cm}p{2.5cm}p{2.5cm}p{2.5cm}}
\hline
\textbf{Malware type} & \textbf{Network} & \textbf{I/O} & \textbf{File system} & \textbf{Resource usage} & \textbf{System call} & \textbf{Kernel events} \\
\hline
Botnets & 
Ilavarasan et al.~\cite{ilavarasan2012} \par 
Meidan et al.~\cite{meidan2018n} \par 
Koroniotis et al.~\cite{koroniotis2019} 
& 
& 
& 
Bezerra et al.~\cite{Bezerra2019IoTDSAO} 
& 
de Costa et al.~\cite{costa2017} \par 
Martinelli et al.~\cite{martinelli2017} \par 
Saracino et al.~\cite{saracino2018} 
& 
Ilavarasan et al.~\cite{ilavarasan2012} \par 
Martinelli et al.~\cite{martinelli2017} 
\\

Backdoors & 
Zhang et al.~\cite{zhang2001} \par 
Huertas et al.~\cite{huertas2022intelligent} 
& 
& 
Huertas et al.~\cite{huertas2022intelligent} 
& 
Huertas et al.~\cite{huertas2022intelligent} 
& 
Canzanese et al.~\cite{canzanese2015} 
& 
Huertas et al.~\cite{huertas2022intelligent} 
\\

Rootkits & 
Hoglund \& Butler~\cite{hoglund2005} 
& 
Kruegel et al.~\cite{kruegel2004} \par 
Baliga et al.~\cite{baliga2008} 
& 
Nick et al.~\cite{nick2004} 
& 
Nick et al.~\cite{nick2004} \par 
Baliga et al.~\cite{arati2011} 
& 
Nick et al.~\cite{nick2004} \par 
Carbone et al.~\cite{carbone2009} 
\\

Ransomware & 
Kok et al.~\cite{kok2019RansomwareT} 
& 
Kok et al.~\cite{kok2019RansomwareT} 
& 
Kok et al.~\cite{kok2019RansomwareT} 
& 
Kok et al.~\cite{kok2019RansomwareT} 
& 
Martinelli et al.~\cite{martinelli2017} \par 
Kok et al.~\cite{kok2019RansomwareT} 
& 
Martinelli et al.~\cite{martinelli2017} \par 
Kok et al.~\cite{kok2019RansomwareT} 
\\

Coinminer & 
Barbhuiya et al.~\cite{barbhuiya2018rads} \par 
Tanana et al.~\cite{tanana2020} 
& 
& 
& 
Barbhuiya et al.~\cite{barbhuiya2018rads} \par 
Tanana et al.~\cite{tanana2020} 
& 
Tanana et al.~\cite{tanana2020} 
& 
Huertas et al.~\cite{huertas2022intelligent} 
\\
\hline
\end{tabular}}
\end{table}

To determine which behavioral dimensions are most relevant for monitoring in IoT malware detection, an analysis was conducted of prior studies on malware families targeting IoT devices. \tablename~\ref{tab:malware_behavior} summarizes how different types of malware, botnets, backdoors, rootkits, ransomware, and coinminers, affect various behavioral sources. As shown in the table, different malware families can influence a range of runtime behaviors, including network activity, input/output operations, file system access, resource usage, system calls, and kernel events. 

Network traffic analysis is consistently leveraged for detecting botnets and backdoors, as shown in studies such as Ilavarasan et al.~\cite{ilavarasan2012} and Huertas et al.~\cite{huertas2022intelligent}, where anomalous communication patterns serve as primary indicators of compromise. Similarly, system call monitoring has been widely adopted for identifying malicious execution behavior~\cite{costa2017, canzanese2015}, particularly in backdoor and botnet scenarios.

For rootkit detection, prior work emphasizes kernel-level monitoring and file system integrity analysis, since rootkits often manipulate low-level operating system components~\cite{nick2004, carbone2009}. Ransomware detection studies frequently rely on file system operations and abnormal input/output behavior due to large-scale file encryption activity~\cite{kok2019RansomwareT}. In contrast, coinminer detection predominantly focuses on anomalous resource consumption patterns such as sustained CPU usage and performance counter deviations~\cite{barbhuiya2018rads, tanana2020}.

These observations indicate that no single behavioral source is sufficient to characterize the diversity of IoT malware activity. Instead, effective detection requires cross-layer visibility spanning communication behavior, execution traces, resource utilization, and file system operations. Therefore, six behavioral dimensions were selected for monitoring: resource usage, kernel events, system calls, network activity, input/output operations, and file system access. Monitoring these dimensions is intended to capture the diverse effects of malware on IoT devices and to enable the construction of a dataset that supports dynamic anomaly detection and malware classification.

\subsection*{Crowdsensing Environment}
The experimental environment was designed to emulate a decentralized IoT crowdsensing scenario in which multiple heterogeneous devices operate concurrently and independently. The testbed was implemented using eight Raspberry Pi devices, comprising six Raspberry Pi 3 and two Raspberry Pi 4 boards. Each device was equipped with a software-defined radio (SDR) kit, serving as the sensing infrastructure. The devices were configured with either 32GB or 64GB SD cards, and ran on an ARM-based CPU architecture using the ElectroSense sensor image to enable data acquisition and processing.

These eight devices were deployed to collect behavioral data under two conditions: a benign (normal) operational state and eight distinct malware attack scenarios. The selected malware covered five major families, botnet, backdoor, rootkit, ransomware, and coinminer. Specifically, the botnet sample used was Bashlite~\cite{bashlite}; the backdoor category included HttpBackdoor~\cite{httpbackdoor}, Backdoor~\cite{backdoor}, and TheTick~\cite{thetick}; the rootkit category included Beurk~\cite{beurk} and Bdvl~\cite{bdvl}; the ransomware sample was Ransomware-PoC~\cite{ransomwarepoc}; and the coinminer sample was XMRig~\cite{xmrig}. All malware samples were obtained from publicly available open-source repositories and executed locally on the devices under controlled conditions.

To simulate sustained malicious activity and ensure consistent behavioral impact, custom execution scripts were developed for each malware sample. These scripts maintained continuous operation by repeatedly triggering representative malicious behaviors, such as file manipulation, network communication, encryption routines, or resource-intensive computation, depending on the malware type. Each experimental session lasted four hours per scenario, resulting in a total monitoring duration of 288 hours across all devices and conditions.

Although the experimental setup focuses on Raspberry Pi devices sharing an ARM-based architecture, the selected hardware platforms represent widely deployed IoT-class devices in edge and sensing applications. The inclusion of two different hardware generations (Raspberry Pi 3 and 4) introduces variability in processing capability and resource characteristics. More importantly, behavioral heterogeneity in this dataset arises primarily from diverse malware families and execution conditions rather than from hardware differences alone. This design choice allows controlled investigation of behavioral artifacts while maintaining reproducibility, and provides a realistic baseline for decentralized learning research in resource-constrained IoT environments.

\subsection*{Device Behavioral Monitoring}
Based on the literature analysis, each device monitored its behavior using six modules, collecting data from the following dimensions: resource usage, kernel events, system calls, network activity, input/output operations, and file system access. Monitoring was performed locally on each device to emulate decentralized data generation in the crowdsensing environment. All behavioral traces were collected continuously during both benign and malware-infected execution.

To ensure temporal consistency across different data sources, system-level metrics were sampled at fixed intervals (primarily every 5 seconds), while certain event-driven logs (e.g., file modifications or packet captures) were recorded continuously and later aligned into uniform time windows during dataset construction. This unified monitoring strategy ensures that heterogeneous behavioral signals can be jointly analyzed while preserving their original granularity.

\subsubsection*{Resource Usage}
This module monitored device-level resource utilization, including CPU and memory usage, disk utilization, network throughput, page faults, cache misses, and hardware performance counters. Metrics were sampled every 5 seconds using standard system utilities and hardware counters exposed via \texttt{perf}. The collected features provided an overview of the system load and bottlenecks during both benign and malicious operations.

\subsubsection*{Kernel Events}
This module recorded fine-grained kernel-level tracepoints at 5-second intervals, using \texttt{perf} to log a predefined set of kernel events indicative of I/O operations, memory management, process scheduling, signal handling, network stack activity, and file system writeback. The monitored events covered a wide range of subsystems, including \texttt{block}, \texttt{jbd2}, \texttt{kmem}, \texttt{sched}, \texttt{writeback}, \texttt{irq}, \texttt{net}, \texttt{signal}, and others. These events provided a low-level view of device behavior during benign and malicious operation, enabling analysis of how malware affects kernel state transitions and resource usage.

\subsubsection*{System Calls}
This module monitored the sequence of system calls executed by processes running on the device. Sampling occurred every 10 seconds by using \texttt{perf}, recording the count and type of system calls observed during the window. The collected data included both user-space initiated calls and kernel-level service calls, enabling analysis of process behavior changes indicative of malware activity. 

\subsubsection*{Network Activity}
This module captured TCP and UDP traffic on the \texttt{eth0} interface of each device using the Python-based \texttt{Scapy} library. For each observed packet, the timestamp, protocol type, source and destination IP addresses, source and destination ports, and packet length were recorded. Data were aggregated in 5-second windows to provide time-resolved network flow characteristics under different scenarios.

\subsubsection*{Input/output Operations}
This module monitored block-level input/output activity and the entropy of modified files. Block activity was recorded using \texttt{iostat}, capturing metrics such as reads, writes, and I/O utilization at 5-second intervals. For each file modification event detected via \texttt{inotifywait}, the Shannon entropy of the first 100 bytes of the file content was calculated as:
\begin{equation}
H = -\sum_{i=1}^{n} p_i \log_2 p_i
\end{equation}
where $p_i$ is the relative frequency of each byte value in the sample. This metric reflects the randomness of file content and is commonly used to detect packed or encrypted malware payloads.

\subsubsection*{File System}
This module logged file system-level operations by recording \texttt{perf} events related to \texttt{ext4}, \texttt{block}, \texttt{jbd2}, and writeback subsystems. Events such as file creation, deletion, modification, and journaling activity were tracked. Data were collected continuously and aggregated into 5-second windows to capture fine-grained changes in file system behavior during normal and malicious execution.

Collectively, these monitoring modules generate heterogeneous raw behavioral logs on each device, and the next stage involves constructing the final dataset from the collected raw behavioral logs.

\subsection*{Dataset Construction}
This subsection describes the procedures applied to process the collected raw behavioral data into a structured and clean dataset suitable for training and evaluation. The processing pipeline includes three main steps: cleansing the data to remove irrelevant, missing, or redundant information; aggregating the raw behavioral records into fixed time-window instances; and transforming the raw monitoring outputs into meaningful statistical features.

Behavioral logs generated on each device were first synchronized based on their timestamps. Continuous event-driven records (e.g., packet captures and file modification events) and periodically sampled metrics (e.g., resource utilization) were aligned into a common temporal reference frame. This integration step ensures that signals originating from different monitoring modules can be jointly analyzed.

\subsubsection*{Data Cleansing}
After integrating the behavioral data from six monitoring modules across eight IoT devices, a data cleansing step was performed to ensure the quality of the dataset and prepare it for subsequent feature selection and model training. A rule-based data cleansing procedure was applied prior to aggregation to remove non-informative or redundant attributes and ensure consistency across devices.
 The cleansing process consisted of the following steps:
\begin{itemize}
    \item \textbf{Elimination of Useless Features}: Metadata fields that do not represent behavioral signals were manually excluded from the feature space. These include raw \texttt{timestamp}, device identifiers, interface names, and connectivity indicators that serve only for logging or indexing. Such fields are necessary for trace alignment but do not contribute to behavioral discrimination and were therefore removed before feature construction.
    \item \textbf{Elimination of Constant Features}: The variance of each candidate feature was computed across the entire dataset. Features exhibiting zero variance (i.e., constant-valued columns) were eliminated, as they provide no discriminative information for downstream analysis. This variance-based filtering step ensures that only informative behavioral attributes are retained in the final dataset.
    \item \textbf{Missing Value Imputation}: Missing entries in the dataset were filled with zeros to maintain a consistent feature space without introducing NaN values.    
\end{itemize}

\subsubsection*{Temporal Aggregation}
To transform fine-grained logs into structured records, all behavioral signals were aggregated into fixed 30-second time windows. This window size represents a trade-off between temporal resolution and statistical stability. From a behavioral modeling perspective, intrusion-related activities in IoT environments, such as scanning bursts, connection flooding, or anomalous protocol exchanges, typically manifest as short-term deviations in traffic intensity and interaction patterns. A 30-second interval is sufficiently fine-grained to capture these transient behavioral shifts without diluting them across overly long observation periods.

At the same time, excessively short windows, such as 1–5 seconds, produce highly sparse feature vectors in resource-constrained IoT settings, where many devices exhibit low baseline activity. Such sparsity increases variance in statistical descriptors and may lead to unstable local model updates in DFL. The 30-second duration ensures that each window contains enough events to compute meaningful aggregate statistics while preserving temporal responsiveness.

Through temporal aggregation, the original 21,582,484 raw behavioral records were consolidated into 342,106 structured time-window instances. This aggregation step preserves event-level information within statistical summaries, enabling the derivation of statistical descriptors while maintaining the temporal dynamics relevant to intrusion detection.

\subsubsection*{Feature Extraction, Encoding, and Selection}
Following temporal aggregation, source-specific encoding strategies were applied to convert heterogeneous behavioral logs into a unified numerical representation. Although some of the monitoring modules already produced encoded metrics (e.g., counts of kernel or resource events per time window), other modules, including input/output operations, network activity, and system calls, required additional processing to extract features.

\begin{itemize}
    \item \textbf{Input/output Operations.} When a file was created or modified, its entropy value was calculated from the first 100 bytes of its content to capture randomness indicative of encryption. To quantify suspicious activity over time, the recorded entropy values were aggregated into 30-second windows. Within each window, the number of files with entropy values greater than or equal to 6 was computed as the feature \texttt{entropy\_file\_count}.    
     \item \textbf{Network Activity.}
    Network traffic was captured at the packet level, recording protocol, source and destination IP addresses and ports, and packet lengths. These data were aggregated into 30-second windows. From each window, 23 features were computed, as shown in \tablename~\ref{tab:network_features}. These features can be grouped into four categories: (i) traffic volume descriptors (e.g., \texttt{PacketCount}, \texttt{TotalLength}, \texttt{AverageBandwidth}), capturing overall communication intensity; (ii) distributional statistics of packet sizes and inter-packet intervals (e.g., \texttt{VarianceLength}, \texttt{MeanInterPacketInterval}), reflecting temporal regularity and burstiness; (iii) protocol and structural indicators (e.g., \texttt{TcpPacketCount}, \texttt{UdpPacketCount}, \texttt{TcpUdpProtocolRatio}), characterizing communication patterns; and (iv) diversity metrics (e.g., \texttt{DifferentSourcePorts}, \texttt{DifferentDestIPs}), representing the heterogeneity of endpoints and connection behavior. These features described both the intensity and structural organization of network activity within each time window, enabling detection of abnormal communication behaviors associated with botnets, backdoors, or data exfiltration.
    \item \textbf{System Calls.} System call traces were collected by monitoring kernel interactions within each 30-second window. The Bag-of-Words (BoW) encoding was applied, producing frequency vectors representing the distribution of system calls.    
\end{itemize}

\begin{table}[h!]
\small
\centering
\caption{Extracted features from network traffic within each 30-second window.}
\label{tab:network_features}
\begin{tabular}{lll}
\hline
\textbf{Network Feature} & \textbf{Description} & \textbf{Data Type} \\
\hline
\texttt{PacketCount} & Number of packets in the window & Integer \\
\texttt{TotalLength} & Sum of packet lengths & Integer \\
\texttt{AverageLength} & Mean packet length & Float \\
\texttt{MedianLength} & Median packet length & Float \\
\texttt{MinLength} & Minimum packet length & Integer \\
\texttt{MaxLength} & Maximum packet length & Integer \\
\texttt{VarianceLength} & Variance of packet lengths & Float \\
\texttt{DifferentSourcePorts} & Number of unique source ports & Integer \\
\texttt{DifferentDestPorts} & Number of unique destination ports & Integer \\
\texttt{TcpPacketCount} & Number of TCP packets & Integer \\
\texttt{UdpPacketCount} & Number of UDP packets & Integer \\
\texttt{TcpUdpProtocolRatio} & Ratio of TCP to UDP packets & Float \\
\texttt{MeanInterPacketInterval} & Mean inter-packet interval & Float \\
\texttt{VarianceInterPacketInterval} & Variance of inter-packet intervals & Float \\
\texttt{MinInterPacketInterval} & Minimum inter-packet interval & Float \\
\texttt{MaxInterPacketInterval} & Maximum inter-packet interval & Float \\
\texttt{FirstDerivativeInterPacketInterval} & First derivative of interval series & Float \\
\texttt{SecondDerivativeInterPacketInterval} & Second derivative of interval series & Float \\
\texttt{AverageBandwidth} & Mean bandwidth consumption & Float \\
\texttt{VarianceBandwidth} & Variance in bandwidth usage & Float \\
\texttt{MinBandwidth} & Minimum bandwidth observed & Float \\
\texttt{MaxBandwidth} & Maximum bandwidth observed & Float \\
\texttt{DifferentDestIPs} & Number of unique destination IPs & Integer \\
\hline
\end{tabular}
\end{table}

After the feature encoding, the dataset contains a comprehensive 687-dimensional feature space covering six behavioral dimensions. While this representation provides rich cross-layer observability, such high dimensionality may impose non-negligible computational and memory overhead on resource-constrained IoT devices participating in decentralized training.

To reflect practical deployment constraints and evaluate lightweight learning scenarios, a reduced subset of 32 representative features was derived for experimental validation. Feature selection was performed using three statistical scoring methods: the Chi-Squared test (chi2), ANOVA F-value, and Mutual Information. Each method evaluates the dependency between individual features and class labels from different statistical perspectives. Features consistently ranked highly across these criteria were retained to ensure both discriminative relevance and dimensional diversity across behavioral sources.

\begin{itemize}
\item \textbf{Chi-Squared Test (\textit{chi2}):} This test computes the chi-squared statistic between each non-negative feature and the class label, identifying features that are strongly dependent on the target classes.
\item \textbf{ANOVA F-value (\textit{f\_classif}):} The ANOVA F-value measures the ratio of variance between classes to the variance within classes. Features with higher F-values are better at distinguishing between classes.
\item \textbf{Mutual Information (\textit{mutual\_info\_classif}):} Mutual information quantifies the dependency between a feature and the target variable using a nonparametric entropy-based estimator. Features with higher scores exhibit stronger association with the labels.
\end{itemize}

\begin{figure}[h!]
    \centering
    \includegraphics[width=1\linewidth]{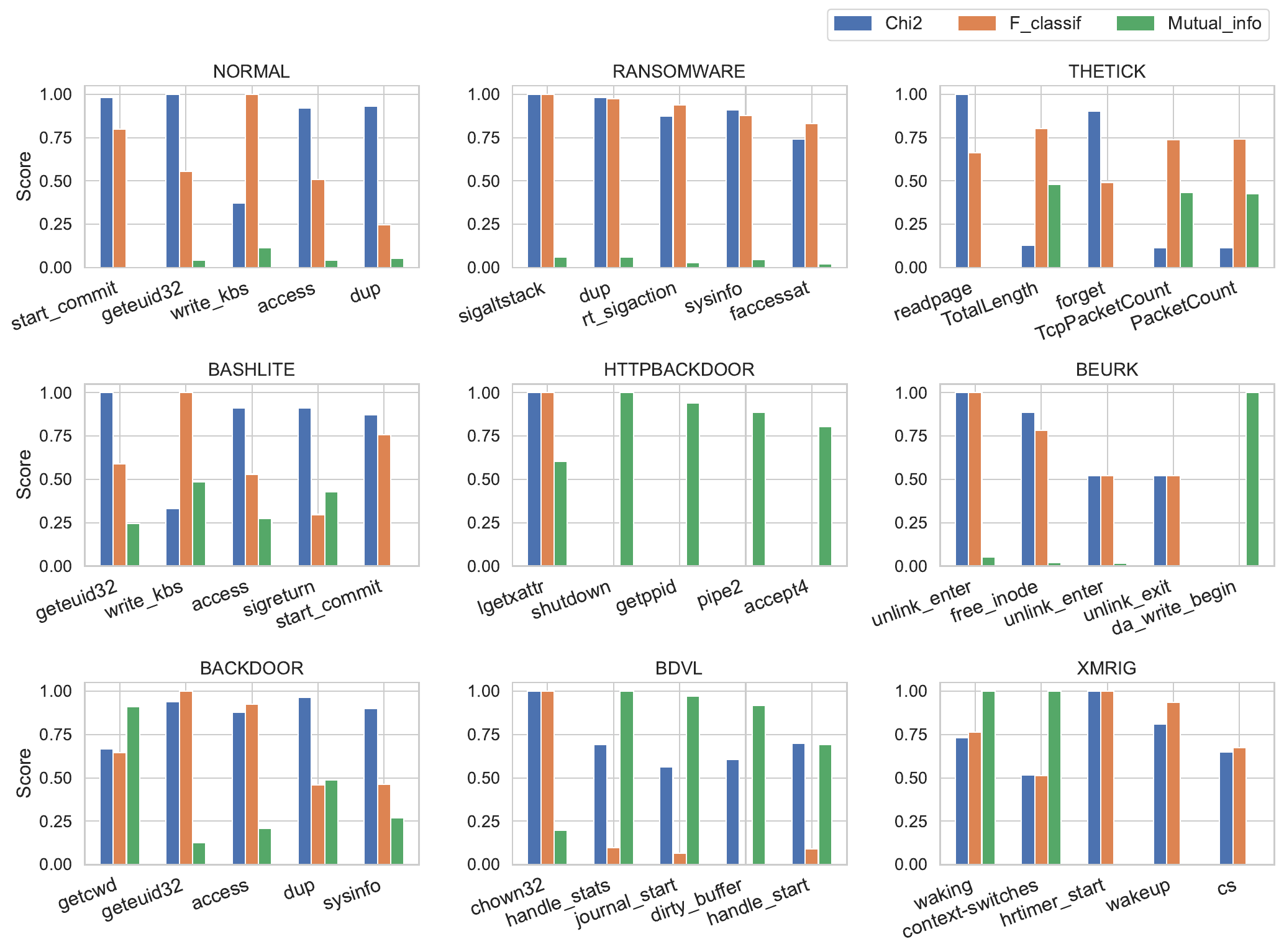}
    \caption{Top-5 features per class ranked by statistical relevance.}
    \label{fig:plot_feature_importance_top_5_per_label}
\end{figure}

To investigate the most relevant features for each class, the top‑5 features per label were selected based on their average scores across the three statistical tests: \textit{Chi‑Squared}, \textit{ANOVA F‑value}, and \textit{Mutual Information}, as shown in \figurename~\ref{fig:plot_feature_importance_top_5_per_label}. The sources of the top features differ across attack categories.

For \textbf{normal traffic}, dominant features include \texttt{start\_commit} and \texttt{getuid32} (System Calls), as well as \texttt{write\_kbs} (Input/Output Operations). These features characterize routine process management and stable I/O activity, representing baseline system behavior without sustained anomalous resource manipulation. In the \textbf{ransomware} category, highly ranked features such as \texttt{sigaltstack}, \texttt{dup}, \texttt{rt\_sigaction}, and \texttt{sysinfo} are all System Calls. These calls are closely related to process control, signal handling, and system state inspection, which aligns with ransomware behavior involving execution context manipulation and system probing prior to encryption routines.

The \textbf{backdoor-related samples}, including \textit{HttpBackdoor}, \textit{Backdoor}, and \textit{TheTick}, exhibit prominent features such as \texttt{shutdown}, \texttt{accept4}, \texttt{getcwd}, \texttt{unlink\_enter}, and \texttt{readpage}. These features originate primarily from File System and Input/Output Operations, reflecting persistent file manipulation, remote command handling, and network-triggered execution. The presence of file unlinking and page-level read operations is consistent with stealth-oriented behaviors often associated with backdoor activity.

For the \textbf{rootkit samples} (\textit{Beurk} and \textit{Bdvl}), influential features include \texttt{unlink\_enter} and \texttt{handle\_stats}, which are associated with System Calls and low-level file system operations. These indicators correspond to kernel-level manipulation and artifact concealment through metadata modification and journaling activity.

In contrast, the \textbf{coinminer sample} (\textit{XMRig}) is characterized by features such as \texttt{waking}, \texttt{context-switches}, and \texttt{wakeup}, which stem from Resource Usage and Kernel Events. These reflect elevated scheduling activity, timer invocation, and CPU wake-up patterns, consistent with sustained computational workloads typical of cryptocurrency mining.

Overall, these observations indicate that different attack families exhibit distinct operational footprints at the system level. Discriminative features are not uniformly distributed across behavioral dimensions but instead concentrate around the functional objectives of each attack type, supporting the dataset's structural validity and semantic coherence. After statistical selection and deduplication, the 32 most important features were identified and used as the input dimensions for the final model training. These feature vectors are normalized using \textit{min–max scaling} to ensure comparability across heterogeneous behavioral metrics.

The resulting dataset contains 342,106 records, each corresponding to a 30-second behavioral snapshot of a device under a specific operational condition. Each record is represented by 32 features spanning six behavioral dimensions and is labeled according to the corresponding benign or malware category. The dataset is stored in CSV format and organized by device and label to facilitate decentralized learning experiments and reproducible evaluation. In addition to the final 32-dimensional feature representation, the dataset also provides the complete raw behavioral logs and the initial 687-dimensional candidate feature space prior to statistical selection, enabling full reproducibility of the feature construction and selection process. The semantic meaning, behavioral dimensions, and representative examples of the 32-dimensional feature set are detailed in the Data Records section.

\section*{Data Records}
This section describes the structure and organization of the released dataset. The dataset consists of 342,106 structured records derived from 21,582,484 raw behavioral logs collected over 288 hours of monitoring. Each record corresponds to a 30-second time window of device activity and contains 32 numerical features spanning six behavioral dimensions, along with a categorical label indicating benign or malware operation.

\subsection*{Dataset Organization}
The complete dataset is released in CSV format and organized hierarchically to reflect the experimental configuration. At the top level, data are partitioned by device, with each of the eight IoT devices assigned a dedicated directory. Within each device directory, records are further grouped by operational condition, including one benign state and eight malware categories (Bashlite, HttpBackdoor, Backdoor, TheTick, Beurk, Bdvl, Ransomware-PoC, and XMRig).

Each CSV file contains structured records corresponding to aggregated 30-second behavioral windows. This organization facilitates decentralized learning experiments by enabling device-level data partitioning without additional preprocessing.

\subsection*{Record Schema}
Each record in the dataset represents a 30-second behavioral snapshot of a single device under a specific operational condition. A record consists of 32 numerical features derived from six behavioral dimensions (resource usage, kernel events, system calls, network activity, input/output operations, and file system access), along with a categorical label indicating the corresponding benign or malware class.

Formally, each record can be represented as: \(\mathbf{x}_i = (f_1, f_2, \dots, f_{32}, y)\), where \( f_j \in \mathbb{R} \) denotes a numerical feature value within the time window, and \( y \) denotes the class label.

All feature columns are numerical and stored as floating-point values after preprocessing. Following data cleansing and temporal aggregation, global min–max normalization was applied to all selected features, scaling their values to the interval $[0,1]$ to ensure numerical comparability across heterogeneous behavioral metrics. As a result, all feature values in the released 32-dimensional representation are bounded within this range. Missing values arising from asynchronous event logging were imputed with zeros prior to normalization to preserve fixed dimensionality across records. After preprocessing and scaling, no NaN values remain in the released dataset. Features exhibiting zero variance across all records were removed before release.

\begin{table*}[h!]
\centering
\small
\caption{Selected 32 features included in the released dataset, their behavioral sources, data type, representative normalized values, and descriptions. All feature values are globally min--max normalized to the interval $[0,1]$.}
\begin{tabular}{lllp{1.8cm}p{5.2cm}}
\hline
\textbf{Feature} & \textbf{Source} & \textbf{Type} & \textbf{Example} & \textbf{Description} \\
\hline
shutdown & System Calls & float & 0.1037 & Process termination system call events. \\
socket & System Calls & float & 0.0413 & Socket creation or manipulation. \\
inotify\_add\_watch & System Calls & float & 0.0003 & Monitor file system changes via inotify. \\
seconds\_RES\_data & Resource Usage & float & 0.0020 & Resource utilization over time. \\
handle\_extend & File System & float & 0.0000 & Journaled file system transaction activity. \\
setgroups32 & System Calls & float & 0.0657 & Set process group IDs. \\
geteuid32 & System Calls & float & 0.0183 & Get effective user ID. \\
pipe2 & System Calls & float & 0.1062 & Create pipe for inter-process communication. \\
da\_update\_reserve\_space\_RES\_data & Resource Usage & float & 0.0000 & Disk space reservation activity. \\
brk & System Calls & float & 0.0373 & Adjust process data segment size. \\
ext\_rm\_leaf & File System & float & 0.0000 & Remove extent leaf in file system. \\
iowritetime & Resource Usage & float & 0.00007 & Time spent writing to I/O devices. \\
ext\_remove\_space\_done & File System & float & 0.0000 & Completed extent space removal. \\
recv & System Calls & float & 0.0306 & Receive data from socket. \\
getegid32 & System Calls & float & 0.0253 & Get effective group ID. \\
iowrite & Resource Usage & float & 0.0248 & Bytes written to I/O devices. \\
prlimit64 & System Calls & float & 0.0389 & Set or get resource limits. \\
statfs64 & System Calls & float & 0.0986 & File system statistics retrieval. \\
fchmod & System Calls & float & 0.0000 & Change file permissions. \\
write\_merge & Resource Usage & float & 0.0280 & Merged write operations. \\
sb\_clear\_inode\_writeback\_KERN\_data & Kernel Events & float & 0.0729 & Kernel inode writeback clearing event. \\
util & Input/Output Events & float & 0.0252 & I/O device utilization metric. \\
write\_kbs & Input/Output Events & float & 0.0388 & Kilobytes written per second. \\
getsockname & System Calls & float & 0.0421 & Retrieve socket name. \\
rename & System Calls & float & 0.0527 & Rename file or directory. \\
block\_unplug\_KERN\_data & Kernel Events & float & 0.0052 & Block device queue unplug event. \\
madvise & System Calls & float & 0.0035 & Advise kernel on memory usage patterns. \\
br\_mis\_pred & Resource Usage & float & 0.2496 & CPU branch misprediction count. \\
setitimer & System Calls & float & 0.2496 & Set interval timer. \\
connect & System Calls & float & 0.0440 & Initiate socket connection. \\
mkdir & System Calls & float & 0.0606 & Create directory. \\
dup2 & System Calls & float & 0.0320 & Duplicate file descriptor. \\
\hline
\end{tabular}
\label{tab:top_32_features}
\end{table*}

\subsection*{Feature Structure and Semantic Characteristics}

The final dataset contains 32 normalized features spanning six behavioral dimensions: System Calls, File System events, Kernel Events, Input/Output operations, and Resource Usage. \tablename~\ref{tab:top_32_features} provides a detailed overview of each feature, including its behavioral source, semantic meaning, and a representative normalized value.

The majority of the selected features originate from System Calls. This reflects the central role of system call activity in characterizing process-level behavior. System calls represent direct interactions between user-space processes and the operating system kernel, and therefore capture fundamental behavioral patterns such as process control, file manipulation, memory management, and network communication. The prominence of these features indicates that variations in system call invocation frequency and composition provide strong discriminative signals for distinguishing benign from malicious activities.

In addition to system calls, several features are derived from Resource Usage and File System events. Resource-related features, such as CPU utilization metrics and I/O throughput measurements, capture quantitative aspects of workload intensity and execution dynamics. These indicators are particularly informative for attacks involving sustained computational activity or abnormal resource consumption. File system features, including journaling and extent modification events, reflect structural changes at the storage layer, which are relevant for behaviors such as persistence mechanisms, data encryption, or artifact manipulation.

Kernel-level events and Input/Output operations further complement the behavioral representation by providing visibility into lower-level system interactions and device-level activity. The selected features form a multi-dimensional representation that captures both control-flow-oriented and resource-oriented characteristics of system behavior. The distribution of features across multiple behavioral dimensions supports the structural completeness of the dataset and enables comprehensive evaluation in decentralized intrusion detection scenarios.

\subsection*{Record Distribution Across Devices and Conditions}

\begin{figure}[h!]
    \centering
    \includegraphics[width=0.75\linewidth]{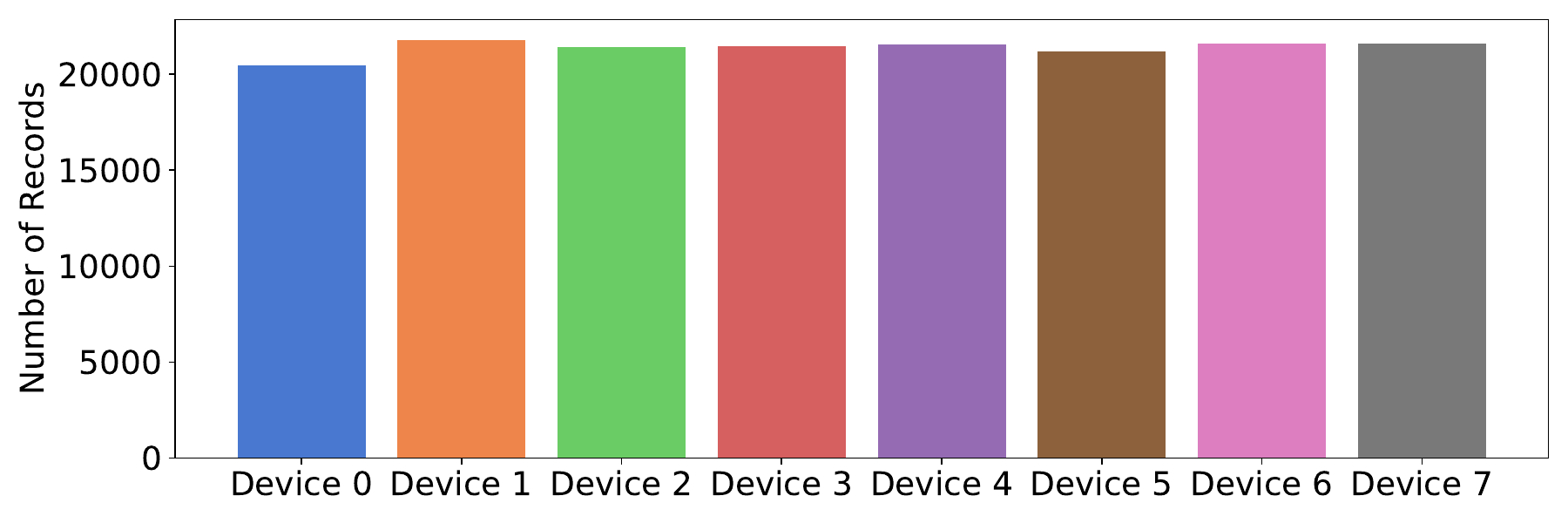}
    \caption{Number of data records per IoT device after feature extraction.}
    \label{fig:device_behavior_counts}
\end{figure}

\figurename~\ref{fig:device_behavior_counts} illustrates the distribution of structured behavioral records across the eight IoT devices after temporal aggregation and feature extraction. The number of records per device is comparable, indicating that the data collection and aggregation process resulted in a balanced distribution across devices. Minor variations arise from execution dynamics and runtime differences during benign and malware scenarios, but no single device dominates the dataset. This balanced distribution supports decentralized learning experiments in which each device can be treated as an independent participant with comparable data volume.

\begin{table}[t]
\centering
\caption{Number of data records per operational condition after feature extraction.}
\label{tab:malware_behavior_counts}
\resizebox{\linewidth}{!}{
\begin{tabular}{lccccccccc}
\hline
 & \textbf{Normal} & \textbf{Ransom-POC} & \textbf{TheTick} & \textbf{Bashlite} & \textbf{HttpBackdoor} & \textbf{Beurk} & \textbf{Backdoor} & \textbf{Bdvl} & \textbf{Xmrig} \\
\hline
\textbf{No. Record} 
& 38,010
& 38,040
& 38,010
& 38,220
& 37,980
& 38,360
& 38,190
& 36,926
& 38,370 \\
\hline
\end{tabular}
}
\end{table}

\tablename~\ref{tab:malware_behavior_counts} reports the number of structured records per operational condition after feature extraction. The dataset includes one benign class and eight malware categories, with each class containing approximately 37,000 to 38,000 aggregated records. This relatively uniform class distribution ensures that the dataset does not exhibit severe class imbalance, thereby enabling fair multi-class classification evaluation under both centralized and decentralized learning settings.

\section*{Technical Validation}
This section validates the usability of the proposed dataset for training DFL models for the proposed IoT crowdsensing intrusion detection dataset. The evaluation focuses on assessing whether the selected 32-dimensional feature representation provides sufficient separability between benign and malicious behaviors under decentralized learning settings.

\subsection*{Experimental Setup}
All experiments were conducted on the Nebular~\cite{beltran2025nebula} platform, which provides a DFL environment with virtualized nodes. The platform supports flexible configurations of datasets, model architectures, data distributions, and network topologies, enabling systematic evaluation of different scenarios.

The experiments were performed using the normalized 32-dimensional feature set. Data were partitioned by device to simulate DFL scenarios, where each device maintains its own local training data. No additional feature engineering or dimensionality reduction was applied beyond the preprocessing steps described earlier. All features were globally min--max normalized to the interval $[0,1]$ prior to training.

As summarized in \tablename~\ref{tab:experiment_config}, each experiment was run for 10 rounds of federated training, with each round consisting of 3 local epochs per node. The experiments compared three approaches: traditional ML, CFL, and DFL. For CFL and DFL, the FedAvg algorithm was used for model aggregation. In terms of DFL, three network topologies, fully connected, ring, and star, were evaluated, with varying numbers of nodes (4, 8, 16, and 32). In the first experiment, an independent and identically distributed (IID) data split was used, while the second experiment employed Dirichlet distributions with \(\alpha=\)10, 1, 0.1 to simulate the impact of non-IID data on the DFL model.

\begin{table}[ht]
\centering
\caption{Experimental configuration overview.}
\label{tab:experiment_config}
\begin{tabular}{ll}
\hline
\textbf{Aspect} & \textbf{Configuration} \\
\hline
Training rounds & 10 \\
Local epochs per round & 3 \\
Model & Three‑layer MLP (\(32 \times 128 \times 9\)) \\
Approaches compared & ML, CFL, DFL \\
DFL topologies & Fully connected, Ring, Star \\
Number of nodes & 4, 8, 16, 32 \\
Data distributions & IID; Non‑IID simulated with Dirichlet (\(\alpha=10, 1, 0.1\)) \\
\hline
\end{tabular}
\end{table}

\subsection*{Baseline Model}

A lightweight multilayer perceptron (MLP) was adopted as the baseline classifier. The model consists of a \(32 \times 128 \times 9\) architecture with ReLU activations. The network architecture was intentionally kept simple to ensure that performance gains could be attributed to the dataset's discriminative structure rather than model complexity. 

The choice of a three-layer MLP is further motivated by deployment considerations in IoT environments. Resource-constrained devices require computationally efficient models with limited memory and processing overhead. The adopted architecture, therefore, balances expressive capacity and computational feasibility, making it suitable for decentralized learning scenarios involving heterogeneous IoT nodes.

Moreover, the experiments were conducted within the Nebula platform using the FedAvg aggregation algorithm. Neural network models are naturally compatible with parameter-based aggregation mechanisms, as model updates can be directly averaged across decentralized participants. Although alternative classifiers such as Random Forests are also powerful, their tree-based structures are not directly compatible with parameter averaging in DFL frameworks. For consistency with the aggregation protocol and to ensure reproducible training, a neural network-based baseline was therefore selected.

\subsection*{Evaluation Metrics}
Model performance was evaluated using standard classification metrics, including accuracy, precision, recall, F1 score, and area under the receiver operating characteristic curve (AUC). For multi-class evaluation, the macro-averaged F1 score was computed to account for class imbalance. Confusion matrices were further analyzed to assess inter-class separability.

To assess communication overhead under different learning paradigms, the total number of transmitted bytes during model update exchange was measured. This metric reflects the network bandwidth consumption associated with centralized and decentralized training settings and provides an estimate of the communication cost introduced by distributed learning.

\subsection*{DFL Model Training and Results}
The first experiment compares different learning paradigms, namely centralized ML, CFL, and DFL, in terms of both model performance and communication overhead. The experimental results are summarized in \tablename~\ref{tab:performance_overhead}. 

\subsubsection*{Model Performance}
When the data can be centralized, as in the traditional ML setting, the model achieves the best performance, with accuracy, F1 score, precision, recall, and AUC all exceeding 0.96 on the complete dataset.

\begin{table}[h!]
\centering
\caption{Performance and communication overhead comparison under different learning scenarios and network sizes.}
\resizebox{\linewidth}{!}{
\begin{tabular}{l r c c c c c c}
\toprule
\textbf{Scenario} & \textbf{Nodes} & \textbf{Accuracy} & \textbf{F1 score} & \textbf{Precision} & \textbf{Recall} & \textbf{AUC} & \textbf{Transmitted Bytes} \\
\midrule
ML & 1 
& 0.963 
& 0.963 
& 0.965 
& 0.963 
& 0.999 
& - \\
\midrule
 & 4  
& 0.942 $\pm$ 0.005 
& 0.942 $\pm$ 0.006 
& 0.943 $\pm$ 0.006 
& 0.942 $\pm$ 0.005 
& 0.998 $\pm$ 0.000 
& 5.117$\times 10^{5}$ $\pm$ 2.621$\times 10^{5}$ \\
CFL & 8  
& 0.898 $\pm$ 0.029 
& 0.897 $\pm$ 0.029 
& 0.900 $\pm$ 0.023 
& 0.898 $\pm$ 0.029 
& 0.993 $\pm$ 0.004 
& 6.746$\times 10^{5}$ $\pm$ 6.460$\times 10^{5}$ \\
 & 16 
& 0.850 $\pm$ 0.032 
& 0.847 $\pm$ 0.033 
& 0.857 $\pm$ 0.025 
& 0.850 $\pm$ 0.032 
& 0.986 $\pm$ 0.005 
& 2.031$\times 10^{6}$ $\pm$ 3.462$\times 10^{6}$ \\
 & 32 
& 0.821 $\pm$ 0.015 
& 0.817 $\pm$ 0.015 
& 0.827 $\pm$ 0.014 
& 0.821 $\pm$ 0.015 
& 0.981 $\pm$ 0.003 
& 4.963$\times 10^{6}$ $\pm$ 1.354$\times 10^{7}$ \\
\midrule
 & 4  
& 0.950 $\pm$ 0.001 
& 0.950 $\pm$ 0.001 
& 0.950 $\pm$ 0.001 
& 0.950 $\pm$ 0.001 
& 0.998 $\pm$ 0.000 
& 1.829$\times 10^{6}$ $\pm$ 6.834$\times 10^{4}$ \\
DFL (Fully) & 8  
& 0.933 $\pm$ 0.002 
& 0.932 $\pm$ 0.002 
& 0.933 $\pm$ 0.002 
& 0.933 $\pm$ 0.002 
& 0.997 $\pm$ 0.000 
& 5.396$\times 10^{6}$ $\pm$ 1.549$\times 10^{5}$ \\
 & 16 
& 0.890 $\pm$ 0.004 
& 0.887 $\pm$ 0.004 
& 0.887 $\pm$ 0.004 
& 0.890 $\pm$ 0.004 
& 0.992 $\pm$ 0.000 
& 2.153$\times 10^{7}$ $\pm$ 3.800$\times 10^{5}$ \\
 & 32 
& 0.914 $\pm$ 0.006 
& 0.913 $\pm$ 0.007 
& 0.916 $\pm$ 0.007 
& 0.914 $\pm$ 0.006 
& 0.995 $\pm$ 0.001 
& 1.854$\times 10^{7}$ $\pm$ 7.816$\times 10^{6}$ \\
\midrule
 & 4  
& 0.946 $\pm$ 0.005 
& 0.946 $\pm$ 0.005 
& 0.947 $\pm$ 0.005 
& 0.946 $\pm$ 0.005 
& 0.998 $\pm$ 0.000 
& 5.147$\times 10^{6}$ $\pm$ 9.002$\times 10^{5}$ \\
DFL (Random) & 8  
& 0.906 $\pm$ 0.009 
& 0.904 $\pm$ 0.010 
& 0.905 $\pm$ 0.010 
& 0.906 $\pm$ 0.009 
& 0.994 $\pm$ 0.001 
& 1.023$\times 10^{7}$ $\pm$ 6.060$\times 10^{6}$ \\
 & 16 
& 0.837 $\pm$ 0.023 
& 0.832 $\pm$ 0.023 
& 0.840 $\pm$ 0.018 
& 0.837 $\pm$ 0.023 
& 0.984 $\pm$ 0.004 
& 3.850$\times 10^{6}$ $\pm$ 1.306$\times 10^{6}$ \\
 & 32 
& 0.864 $\pm$ 0.026 
& 0.861 $\pm$ 0.027 
& 0.869 $\pm$ 0.025 
& 0.864 $\pm$ 0.026 
& 0.989 $\pm$ 0.004 
& 1.760$\times 10^{7}$ $\pm$ 8.187$\times 10^{6}$ \\
\midrule
 & 4  
& 0.951 $\pm$ 0.001 
& 0.951 $\pm$ 0.001 
& 0.951 $\pm$ 0.000 
& 0.951 $\pm$ 0.001 
& 0.998 $\pm$ 0.000 
& 1.204$\times 10^{6}$ $\pm$ 3.702$\times 10^{4}$ \\
DFL (Ring) & 8  
& 0.928 $\pm$ 0.002 
& 0.927 $\pm$ 0.002 
& 0.928 $\pm$ 0.003 
& 0.928 $\pm$ 0.002 
& 0.997 $\pm$ 0.000 
& 1.317$\times 10^{6}$ $\pm$ 6.384$\times 10^{4}$ \\
 & 16 
& 0.825 $\pm$ 0.019 
& 0.820 $\pm$ 0.019 
& 0.830 $\pm$ 0.019 
& 0.825 $\pm$ 0.019 
& 0.981 $\pm$ 0.003 
& 7.789$\times 10^{6}$ $\pm$ 3.249$\times 10^{6}$ \\
 & 32 
& 0.874 $\pm$ 0.029 
& 0.871 $\pm$ 0.030 
& 0.877 $\pm$ 0.028 
& 0.874 $\pm$ 0.029 
& 0.990 $\pm$ 0.004 
& 1.013$\times 10^{7}$ $\pm$ 3.014$\times 10^{6}$ \\
\midrule
 & 4  
& 0.938 $\pm$ 0.004 
& 0.938 $\pm$ 0.004 
& 0.939 $\pm$ 0.004 
& 0.938 $\pm$ 0.004 
& 0.997 $\pm$ 0.000 
& 3.342$\times 10^{6}$ $\pm$ 2.147$\times 10^{6}$ \\
DFL (Star) & 8  
& 0.905 $\pm$ 0.007 
& 0.904 $\pm$ 0.007 
& 0.906 $\pm$ 0.007 
& 0.905 $\pm$ 0.007 
& 0.994 $\pm$ 0.001 
& 9.941$\times 10^{6}$ $\pm$ 1.280$\times 10^{7}$ \\
& 16 
& 0.872 $\pm$ 0.008 
& 0.867 $\pm$ 0.009 
& 0.872 $\pm$ 0.008 
& 0.872 $\pm$ 0.008 
& 0.989 $\pm$ 0.001 
& 1.023$\times 10^{7}$ $\pm$ 1.846$\times 10^{7}$ \\
 & 32 
& 0.911 $\pm$ 0.010 
& 0.910 $\pm$ 0.010 
& 0.913 $\pm$ 0.010 
& 0.911 $\pm$ 0.010 
& 0.995 $\pm$ 0.001 
& 8.519$\times 10^{6}$ $\pm$ 2.200$\times 10^{7}$ \\
\bottomrule
\end{tabular}
}

\label{tab:performance_overhead}
\end{table}

The DFL approach preserves data privacy by keeping data local to each node, but at the cost of a slight drop in model performance. For example, with four nodes and a fully connected topology, the DFL model achieves an F1 score of approximately 0.95, which is lower than ML but still competitive. This performance drop is mainly attributed to the reduced amount of data available in each node. DFL achieves comparable or superior performance to CFL under most evaluated configurations, indicating that the dataset characteristics and heterogeneity are better suited to the decentralized training paradigm proposed in this work.

In terms of scalability, the results show that increasing the number of nodes negatively impacts model performance. As the number of nodes increases from 4 to 32, the F1 score decreases, reflecting the diminishing amount of data allocated to each node and the increasing challenge of maintaining global consistency.

Regarding the impact of network topology, the results suggest that the choice of topology has only a minor effect on model performance for this dataset. Fully connected, ring, and star topologies yield comparable results for the same node count, indicating that the dataset is insensitive to the DFL network's communication topology.

\subsubsection*{Network Overhead}
In addition, the impact of network topology on communication overhead in DFL is analyzed. The results show that the total amount of transmitted data in DFL generally increases with the number of participating nodes, while it is also strongly influenced by the underlying network topology. Denser topologies, such as fully connected graphs, lead to higher communication volumes due to more frequent and redundant model exchanges among nodes. In contrast, sparser topologies, such as the ring or star, incur lower communication costs as information propagation is constrained by fewer communication links. These observations highlight an inherent trade-off in DFL between model performance and network communication overhead, where improved collaboration and information sharing come at the cost of increased communication.

\subsubsection*{Convergence Dynamics}
\begin{figure}[h!]
    \centering
    \includegraphics[width=0.5\linewidth]{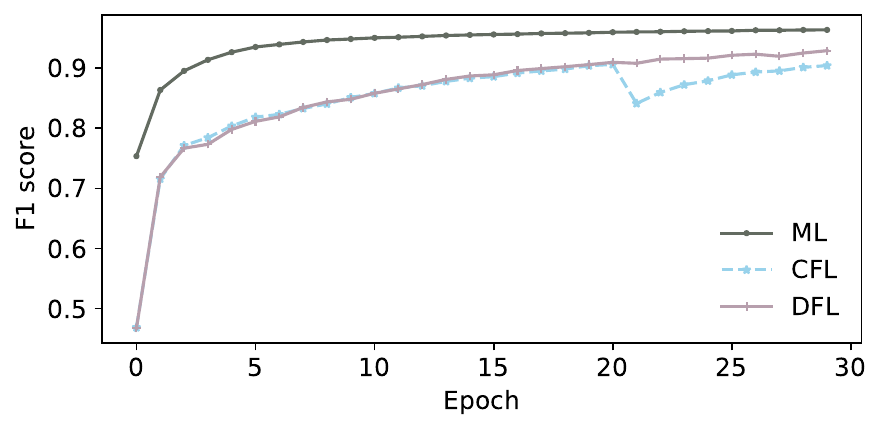}
    \caption{Training F1 score across training epochs for centralized ML, CFL (8 nodes), and DFL (8 nodes, fully connected topology) under IID data distribution.}
    \label{fig:convergence}
\end{figure}

To further analyze convergence dynamics across learning paradigms, \figurename~\ref{fig:convergence} illustrates the evolution of the macro F1 score over training epochs under IID data distribution. Centralized ML exhibits the fastest convergence and achieves the highest final performance. This behavior is attributed to full data exposure during training, enabling direct optimization over the complete dataset without distributed aggregation constraints.

Both CFL and DFL demonstrate slower convergence in the early epochs due to decentralized data partitioning and iterative model aggregation. However, their convergence trajectories remain stable and closely aligned throughout training. DFL achieves slightly higher final performance than CFL in this configuration. This result is consistent with the overall performance trends reported in \tablename~\ref{tab:performance_overhead} and suggests that the dataset's structural characteristics, particularly device-level behavioral consistency and cross-layer feature coherence, are well-suited to decentralized peer-to-peer aggregation. Overall, the convergence analysis confirms that the proposed dataset supports stable distributed optimization while preserving competitive classification performance under decentralized training.

\begin{figure}[b]
    \centering
    \includegraphics[width=0.65\linewidth]{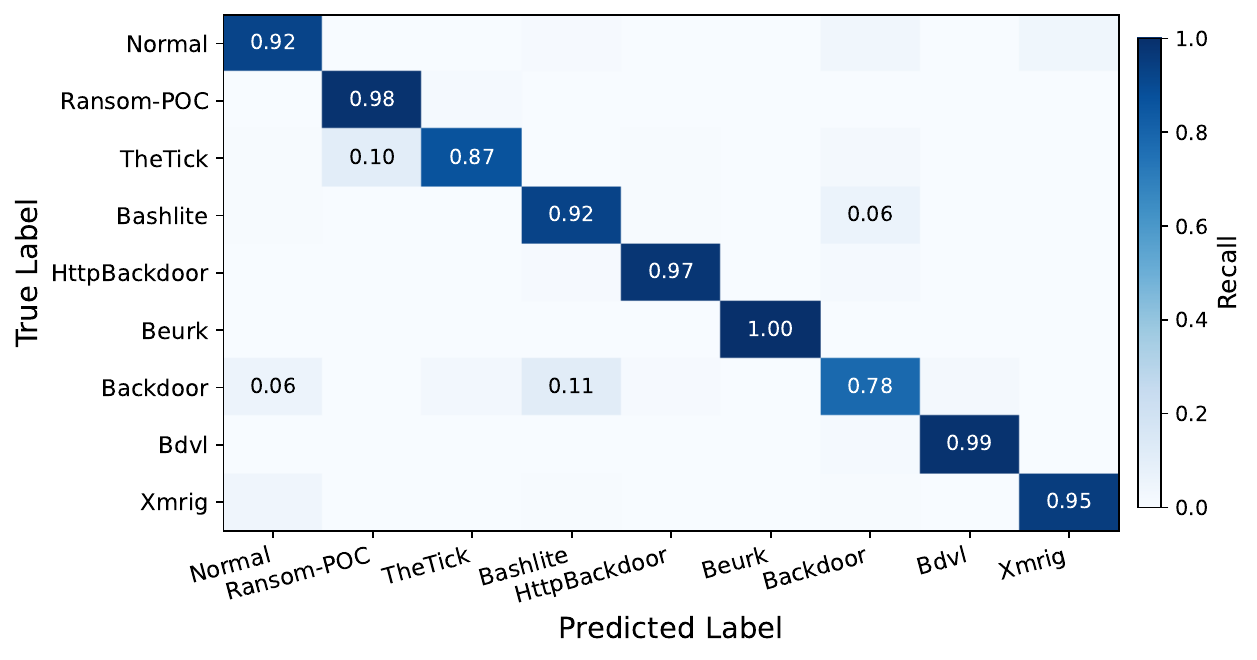}
    \caption{Normalized confusion matrix of the DFL fully connected topology with 8 nodes under IID data distribution.}
    \label{fig:cm_dfl_8_iid}
\end{figure}

\figurename~\ref{fig:cm_dfl_8_iid} presents the normalized confusion matrix for the DFL of a fully connected topology with 8 participating nodes under the IID data distribution. Most categories exhibit strong diagonal dominance, indicating high per-class recall. In particular, Beurk and Bdvl are classified with near-perfect accuracy, while Ransomware-PoC and HttpBackdoor also demonstrate high separability from other classes.  However, certain malware categories present greater classification difficulty. The Backdoor class achieves a recall of 0.78, with a portion of samples misclassified as Bashlite and, to a lesser extent, as Normal traffic. This suggests partial behavioral overlap in network or execution patterns between these classes. Similarly, TheTick exhibits moderate confusion with Ransomware-PoC, reflected by a recall of 0.87. These misclassifications indicate that some malware families share overlapping behavioral artifacts within aggregated time windows. Overall, the confusion matrix confirms that the constructed feature space enables effective multi-class separation, while also revealing subtle similarities between specific malware behaviors under decentralized training.

\subsubsection*{Data Heterogeneity}
To analyze the impact of data heterogeneity on decentralized training performance, experiments were conducted under different Dirichlet distribution parameters (\(\alpha = 10, 1, 0.1\)) in an 8-node fully connected DFL setup. Larger values of \(\alpha\) correspond to more homogeneous data distributions, while smaller values induce stronger heterogeneity across devices.

As shown in \tablename~\ref{tab:dfl_noniid}, performance remains stable when transitioning from near-IID conditions (\(\alpha = 10\)) to moderately non-IID settings (\(\alpha = 1\)), with accuracy and macro F1 scores remaining around 0.93. However, under highly heterogeneous conditions (\(\alpha = 0.1\)), performance degrades more noticeably, with accuracy decreasing to 0.813 and the macro F1 score to 0.807. However, the total number of transmitted bytes remains similar across different Dirichlet settings, indicating that data heterogeneity primarily affects model performance rather than communication overhead in the DFL setup.

\begin{table}[h!]
\centering
\small
\caption{DFL model performance with a fully connected topology with 8 nodes in non-IID setting.}
\begin{tabular}{l c c c c c c}
\toprule
\textbf{Dirichlet $\alpha$} &  \textbf{Accuracy} & \textbf{F1 score} & \textbf{Precision} & \textbf{Recall} & \textbf{AUC} & \textbf{Transmitted Bytes} \\
\midrule
10 
& 0.931 $\pm$ 0.002
& 0.930 $\pm$ 0.003
& 0.930 $\pm$ 0.003
& 0.931 $\pm$ 0.002
& 0.997 $\pm$ 0.000
& 5.648$\times 10^{6}$ $\pm$ 1.513$\times 10^{5}$ \\
1
& 0.930 $\pm$ 0.002
& 0.930 $\pm$ 0.001
& 0.931 $\pm$ 0.001
& 0.930 $\pm$ 0.002
& 0.997 $\pm$ 0.000
& 5.684$\times 10^{6}$ $\pm$ 9.804$\times 10^{4}$ \\
0.1 
& 0.813 $\pm$ 0.005
& 0.807 $\pm$ 0.007
& 0.858 $\pm$ 0.013
& 0.813 $\pm$ 0.005
& 0.986 $\pm$ 0.000
& 5.732$\times 10^{6}$ $\pm$ 5.734$\times 10^{4}$ \\
\bottomrule
\end{tabular}
\label{tab:dfl_noniid}
\end{table}

Despite this degradation, the AUC remains high (0.986), suggesting that the feature space retains discriminative structure even under severe data heterogeneity. These results indicate that the constructed dataset supports stable decentralized training under moderate distribution shifts, while highlighting the challenges posed by extreme non-IID (low \(\alpha\)) scenarios.

\subsubsection*{Robustness for Adversarial Attacks}
To evaluate the robustness of decentralized training under adversarial conditions, label flipping attacks were introduced by randomly selecting a proportion of participating nodes and inverting their local labels~\cite{feng2024dart}. \tablename~\ref{tab:dfl_poisoning} reports the performance degradation under poisoning ratios of 0.25, 0.50, and 0.75 in an 8-node fully connected DFL topology.

\begin{table}[h!]
\centering
\small
\caption{Impact of data poisoning attacks with different poisoning ratios on model performance in the DFL with a fully connected topology with 8 nodes.}
\begin{tabular}{c c c c c c c}
\toprule
\textbf{Poisoning Ratio}& \textbf{Accuracy} & \textbf{F1 score} & \textbf{Precision} & \textbf{Recall} & \textbf{AUC} \\
\midrule
0.25
& 0.853 $\pm$ 0.019
& 0.843 $\pm$ 0.022
& 0.876 $\pm$ 0.013
& 0.853 $\pm$ 0.019
& 0.990 $\pm$ 0.002 \\
0.50
& 0.360 $\pm$ 0.044
& 0.305 $\pm$ 0.043
& 0.356 $\pm$ 0.089
& 0.360 $\pm$ 0.044
& 0.856 $\pm$ 0.029 \\
0.75
& 0.108 $\pm$ 0.013
& 0.100 $\pm$ 0.011
& 0.095 $\pm$ 0.009
& 0.108 $\pm$ 0.013
& 0.631 $\pm$ 0.010\\
\bottomrule
\end{tabular}

\label{tab:dfl_poisoning}
\end{table}

When 25\% of nodes are adversarial, the model maintains relatively stable performance (Accuracy = 0.853, F1 = 0.843), indicating that the decentralized aggregation mechanism tolerates a limited fraction of corrupted participants. However, when the poisoning ratio increases to 50\%, performance drops substantially (Accuracy = 0.360, F1 = 0.305), suggesting that the learning process becomes unstable as malicious updates dominate aggregation. Under extreme conditions (75\% adversarial nodes), performance approaches near-random levels (Accuracy = 0.108, F1 = 0.100), demonstrating that the decentralized setup exhibits substantial performance degradation under large-scale label manipulation. These results highlight both the vulnerability of DFL under high adversarial participation and the importance of incorporating robust aggregation or defense mechanisms in practical deployments.

The validation experiments indicate that the collected dataset supports stable training under DFL settings while preserving data locality. Comparable performance trends are observed across centralized and decentralized configurations, confirming that the extracted feature space retains discriminative behavioral information. The results further show that increasing the number of participating nodes or introducing stronger non-IID data distributions leads to performance degradation, reflecting the impact of statistical heterogeneity and distributed data fragmentation. These observations are consistent with known characteristics of decentralized learning systems. Overall, the dataset provides a controlled but realistic benchmark for evaluating privacy-preserving and distributed learning approaches under varying conditions of scale, distribution skew, and adversarial participation.

\subsection*{Limitations and Future Work}
Several limitations of the current dataset and validation setup should be acknowledged. First, the experimental platform focuses on ARM-based Raspberry Pi devices, which represent a common class of IoT hardware but do not cover the full diversity of real-world IoT ecosystems. Devices with substantially different operating systems, sensing capabilities, or resource profiles may exhibit additional behavioral variability not captured in this release.

Second, temporal aggregation was performed using a fixed 30-second window. While this interval balances temporal resolution and statistical stability, alternative window sizes may reveal finer-grained or longer-term behavioral dynamics. Exploring adaptive or multi-scale aggregation strategies remains an avenue for future investigation.

Third, adversarial evaluation was limited to untargeted label flipping attacks under controlled conditions. More sophisticated poisoning strategies, Byzantine behaviors, or adaptive attackers were not considered in this study. Incorporating robust aggregation mechanisms and evaluating stronger adversarial models would further extend the applicability of the dataset.

Future extensions of this dataset may include additional malware families, expanded benign activity profiles, and deployment across more heterogeneous hardware platforms. Such extensions would further enhance its suitability as a benchmark for distributed and privacy-preserving learning research in IoT environments.

\section*{Usage Notes}
The dataset supports research on DFL, distributed intrusion detection, and behavioral malware classification in IoT environments. The released materials include three levels of data representation: (i) raw behavioral logs collected from each device, (ii) the fully processed 687-dimensional feature dataset after cleansing, aggregation, and encoding, and (iii) a reduced 32-dimensional feature subset used for lightweight validation experiments under resource-constrained settings.

Researchers may choose the appropriate representation depending on their objectives. The full 687-dimensional feature space preserves comprehensive cross-layer behavioral information and enables alternative feature selection, dimensionality reduction, or representation learning approaches. The 32-feature subset is provided to facilitate reproducible baseline experiments in resource-aware or decentralized scenarios.

Although the aggregated dataset exhibits relatively balanced class distributions, decentralized learning experiments may introduce statistical heterogeneity depending on the chosen partitioning strategy. Users are therefore encouraged to explicitly define IID or non-IID partitioning schemes when evaluating distributed learning methods.

For supervised learning tasks, standard preprocessing steps such as feature normalization or scaling are recommended. Stratified data splitting should be applied in centralized experiments to preserve label proportions. When conducting federated or decentralized evaluations, reporting communication configurations and data partition strategies is essential for reproducibility. The dataset can further serve as a benchmark for studying adversarial robustness, communication-efficient aggregation, and lightweight model design in IoT environments.

\section*{Data Availability}
The complete dataset is available from the Science Data Bank~\cite{feng2025iot}, comprising CSV files organized by device and label, with each file containing preprocessed behavioral records and the corresponding extracted features.

\section*{Code Availability}
The scripts for data collection are available at:
\url{https://github.com/Cyber-Tracer/MalwareDetectionDataset}
and the scripts for data processing and model training are provided at:
\url{https://github.com/Cyber-Tracer/iot-feature-engineering}.

\section*{Acknowledgements}
This work was supported by the Swiss Federal Office for Defense Procurement (armasuisse) under the CyberDFL project (CYD-C-2020003) and by the University of Zürich UZH.

\section*{Author contributions statement}
C.F. and A.H.C. drafted the manuscript with contributions from all authors. J.H., H.Q., X.C., Z.Z., and L.K. performed the experiments and data analysis. G.B. conceived and designed the study. B.S. provided overall supervision. All authors reviewed and approved the final manuscript.

\section*{Ethics declarations}
\subsection*{Competing interests}
The authors declare no competing interests.

\end{document}